\documentclass[12pt]{emulateapj} 
\usepackage{amsmath}
\usepackage{amssymb}
\usepackage{amstext}
\usepackage{graphicx}
\usepackage{epsfig}
\usepackage{verbatim}
\usepackage{color}
\definecolor{myColor}{rgb}{0.9,0.9,0.9}    
\begin{document}
\renewcommand\bottomfraction{.9}
\shorttitle{Thermal inversions in hot Jupiter atmospheres}
\title{On the inference of thermal inversions in hot Jupiter atmospheres.}
\author{N. Madhusudhan\altaffilmark{1}$^{,}$\altaffilmark{3} \& 
  S. Seager\altaffilmark{1}$^{,}$\altaffilmark{2}}
\altaffiltext{1}{MIT Kavli Institute for Astrophysics and Space Research, and 
  Department of Earth, Atmospheric, and Planetary Sciences, MIT, Cambridge, MA, 02139}
\altaffiltext{2}{Department of Physics, MIT, Cambridge, MA 02139} 
\altaffiltext{3}{Corresponding author:{\tt nmadhu@mit.edu}} 

\begin{abstract}
  
Several studies in the recent past have inferred the existence of 
thermal inversions in some transiting hot Jupiter atmospheres. Given the limited data 
available, the inference of a thermal inversion depends critically on the chemical 
composition assumed for the atmosphere. In this study, we explore the degeneracies 
between thermal inversions and molecular abundances in four highly irradiated hot Jupiter 
atmospheres, day-side observations of which were previously reported to be 
consistent with thermal inversions based on {\it Spitzer} photometry. The four systems 
are: HD~209458b, HAT-P-7b, TrES-4, and TrES-2. 
We model the exoplanet atmospheres using a 1-D line-by-line radiative transfer code with 
parametrized abundances and temperature structure, and with constraints of energy balance 
and hydrostatic equilibrium. For each system, we explore the model parameter space with 
$\sim 10^6$ models using a Markov chain Monte Carlo routine. Our results primarily suggest that a thorough 
exploration of the model parameter space is necessary to identify thermal inversions 
in hot Jupiter atmospheres. We find that existing observations of TrES-4 and TrES-2 
can both be fit very precisely with models with and without thermal inversions, and with 
a wide range in chemical composition. On the other hand, observations of HD~209458b 
and HAT-P-7b are better fit with thermal inversions than without, as has been reported 
previously. Physically plausible non-inversion models of HD~209458b and HAT-P-7b fit the 
data only at the 1.7$\sigma$ observational errors; better fits require substantial enhancement of methane 
and depletion of CO, which seems implausible in the very hot atmospheres considered here. Secondly, 
in the sample under consideration here, we do not see a correlation between irradiation levels and 
thermal inversions, given current data. Before JWST becomes available, near-IR observations from 
ground and with HST, along with existing {\it Spitzer} observations, can potentially resolve thermal 
inversions in some systems. Observations with only two channels of {\it Warm Spitzer} photometry 
and good S/N can likely identify or rule out thermal inversions if the difference between the fluxes 
in the 3.6 and 4.5 $\micron$ channels is very high.

\end{abstract}

\keywords{planetary systems --- planets and satellites: general  --- 
  radiative transfer}

\section{Introduction}
\label{sec:intro}

Observations of day-side atmospheres of several hot Jupiters have 
indicated the existence of thermal inversions. The inference of
thermal inversions are a result of high S/N {\it Spitzer} observations
(Knutson et al. 2008; Knutson et al. 2009; Machalek et al. 2009; O'Donovan et al. 2010;
Christiansen et al. 2010) and concomitant theoretical modeling (Burrows
et al.  2007, 2008; Fortney et al. 2008; Madhusudhan \& Seager, 2009). 
Thermal inversions have been reported based on flux excesses in 
certain {\it Spitzer} channels. A natural explanation of channel-specific high 
flux is to invoke molecular emission features, as opposed to absorption 
features (Burrows et al. 2007, 2008; Fortney et al. 2008). Under the 
assumption of local thermodynamic equilibrium (LTE), molecular 
emission features form only in the presence of a thermal inversion, 
a region in the atmosphere where temperature increases outwards, 
as in the Earth's stratosphere.

Preceding recent observations, early theoretical work on hot Jupiter 
atmospheres using self-consistent 1D atmosphere models predicted 
the existence of thermal inversions based on absorption due to TiO 
and VO (Hubeny et al. 2003, Fortney et al. 2006). More recently, 
Fortney et al. 2008, classified hot Jupiters in two categories based on 
the degree of incident irradiation. The class of atmospheres with very 
high incident flux, dubbed ``very hot Jupiters'' or ``pM'' class, were 
considered favorable to host thermal inversions caused due to gaseous 
TiO and VO absorption at low pressures, and atmospheres with lower 
fluxes were predicted to be unlikely to host inversions owing to condensation 
of TiO/VO. The flux boundary for this dichotomy was, somewhat arbitrarily, 
chosen to be $\sim 10^9$ ergs/s/cm$^2$. However, while inferences of some 
recent observations have purported to violate this hypothesis 
(e.g. Machalek et al. 2009), others have found present observations 
insufficient to test this hypothesis (O'Donovan et al. 2010; Fressin et al. 2010).

The theory behind the absorbers causing thermal inversions in hot Jupiters 
atmospheres is still under debate; discussed in detail in \ref{sec:theory}. Recent 
theoretical studies suggest that TiO/VO may not be able to totally account for the 
inferred thermal inversions (Spiegel et al. 2009). Other alternatives proposed in 
recent studies include strong UV/visible absorption due of photochemically 
produced sulfur compounds (Zahnle et al. 2009), and correlation of thermal inversions 
with chromospheric activity of host stars (Knutson et al. 2010). Since the nature of 
absorbers causing the inversions is not known, models that have been successful 
in inferring thermal inversions have either adopted a parametric absorber (Burrows et al. 2008), 
or a parametrized temperature profile (see e.g. Madhusudhan \& Seager, 2009). 

Existing models inferring thermal inversions rest on several assumptions and 
parametrization. Models typically invoke several free parameters to induce a thermal inversion 
in the temperature profile, as suggested by the data. The free parameters range 
from the location and magnitude of an unknown opacity source (e.g. Burrows et al. 2007 
\& 2008) to parametrizing the temperature profile itself (e.g. Madhusudhan \& Seager, 2009). 
In addition, traditionally exoplanet atmosphere models span only a limited range in 
chemical abundances, often assuming thermochemical equilibrium (Barman et al. 2005; 
Seager et al. 2005; Fortney et al. 2006; Burrows et al. 2007 \& 2008). While some reported 
models require thermal inversions to explain the observations, it is not known if the model 
parameter space has been thoroughly exhausted. It remains to be seen whether some of 
the observations can be explained without thermal inversions, if we were to relax many 
of the assumptions and thoroughly explore the parameter space. 

In the present work, our primary goal is to understand the extent to which thermal inversions 
can be robustly inferred in hot Jupiter atmospheres, with current observations. We choose a 
test sample of four systems for which {\it Spitzer} observations in the past have been reported 
to be consistent with thermal inversions. We then pursue a detailed exploration of the model 
parameter space to see the extent to which the observations can be explained by models with 
and without thermal inversions. We accomplish this by computing large ensembles of inversion 
and non-inversion models ($N \sim 10^6$), exploring the parameter space for best-fitting solutions.  
For each system, we report quantitatively how well the data can be fit by models with and without thermal 
inversions and with what ranges in atmospheric chemical composition. This approach also reveals 
the underlying correlations between the different chemical species and between the 
composition and temperature structure. And, since the systems considered here have 
different levels of irradiation, we also seek to understand if the presence or absence of 
thermal inversions is correlated with the degree of irradiation, at the level of current data.

We focus on four hot Jupiters for which existing observations were reported to be 
consistent with thermal inversions in their day-side atmospheres, and for which photometric 
observations of thermal emission are available in four or more {\it Spitzer} channels. 
The planets are: HD~209458b, 
HAT-P-7b, TrES-4, and TrES-2. Burrows et al. (2008) and Knutson et al. (2008) first reported a 
thermal inversion in HD~209458b, based on {\it Spitzer} photometry in five channels reported by 
Knutson et al. (2008) and Deming et al. (2005). Madhusudhan \& Seager (2009) confirmed 
the thermal inversion in HD~209458b, for model fits within the 1.4$\sigma$ observational uncertainties. 
Additional observations of HD~209458b are available in the near-IR, obtained 
with {\it HST} NICMOS, but were explained by models with and without inversions 
alike (Swain et al. 2009). Knutson et al. (2009) reported a thermal inversion in 
TrES-4b based on five-channel {\it Spitzer} photometry using models based on 
Burrows et al. (2008). O'Donovan et al. (2010) reported observations of 
TrES-2 in four {\it Spitzer} IRAC Channels, which were explained by models 
with and without thermal inversions; they preferred the inversion model which 
seemed more favorable amongst the set of models explored in that study. 
Croll et al. (2010) reported a ground based detection of TrES-2 in the Ks band, and noted 
that their observation along with the four {\it Spitzer} observations could be 
explained equally well by models with and without thermal inversions. More recently, 
however, Spiegel \& Burrows (2010) reported that models assuming radiative and 
chemical equilibrium cannot explain the {\it Spitzer} observations of TrES-2 without 
invoking thermal inversions. And, finally, Christiansen et al. (2010) reported a thermal 
inversion in HAT-P-7b, based on four {\it Spitzer} IRAC channels, along with an 
observation in the visible by the {\it Kepler Space Telescope} (Borucki et al. 2009). While the Kepler 
point was not necessarily decisive in constraining the thermal inversion itself, 
it allowed constraints on the albedo, day-night redistribution and the TiO/VO 
abundance. Spiegel \& Burrows (2010) confirmed the presence of a thermal 
inversion in HAT-P-7b using the observations of Christiansen et al. (2010). 

The observations of TrES-2 and HAT-P-7b noted above were first interpreted using the 
exoplanet atmosphere model developed in Madhusudhan \& Seager (2009), which is also 
used in the present work. In both those studies, i.e of O'Donovan et al. (2010) and 
Christiansen et al. (2010), we had reported a representative set of models which explained 
the data. In the present study, we report a more exhaustive exploration of the model parameter 
space  for these systems using a new parameter space exploration scheme described in 
Section~\ref{sec:model-mcmc}.

We discuss the theoretical and observational basis of thermal
inversions in \S~2. We explain the model framework in \S~3, along with
the parameter exploration method and the selection of systems for our
study. In \S~4, we present our results, followed by a summary and
discussion in \S~5.

\section{Thermal inversions in hot Jupiter atmospheres}

Several arguments have been proposed in the literature justifying the
existence of thermal inversions in some hot Jupiter atmospheres.
Compelling {\it Spitzer} IRAC observations of some hot Jupiters
suggest that anomalies in the form of planet-star flux excess in some
channels cannot be explained without invoking thermal inversions
(Burrows et al. 2007; Knutson et al. 2008; Madhusudhan \& Seager,
2009; Swain et al. 2009). However, it is not known whether such an
inference is an outcome of the model input choices used to infer the
observations.  On the other hand, independent of observations, several
theoretical arguments support the existence of thermal inversions in
some hot Jupiter atmospheres (Hubeny et al. 2003; Burrows et al. 2007;
Fortney et al. 2008).

In this section, we explore the arguments in favor of thermal inversions, 
and motivate the framework under which they can be tested. We begin with 
the theoretical motivation for thermal inversions, followed by arguments 
leading to inference of thermal inversions from observations. We then pose 
the question of whether the observations can be explained without thermal 
inversions if some of the model constraints are relaxed. Finally, we set up 
the framework in which the requirement of thermal inversions can be 
robustly tested. 

\subsection{Theoretical Basis for Thermal Inversions}
\label{sec:theory}
Thermal inversions are a natural consequence of visible/UV absorption
of incident star light high in the atmosphere. For an isolated planetary
atmosphere in hydrostatic equilibrium, and no local energy sources, the 
atmospheric temperature decreases with pressure (i.e with increasing distance from center); 
the atmosphere is heated from below, and cools monotonically outwards. 
However, in planetary atmospheres irradiated by the host 
star, strong optical/UV absorbers in the higher layers of the atmosphere 
can intercept part of the incident star light. Such local deposition of
energy results in a zone in the planetary atmosphere where temperature
increases outward, i.e a ``thermal inversion''. Most solar system planets have thermal
inversions in their atmospheres. In Earth's atmosphere, for example, a 
thermal inversion is caused by ozone (O$_3$), which is a strong absorber in the UV
(Chamberlain, 1978). And, in Jupiter's atmosphere, a thermal inversion
is caused by strong absorption in the visible by haze resulting from
methane photochemistry.

\begin{figure}[h]
\centering
\includegraphics[width = 0.5\textwidth]{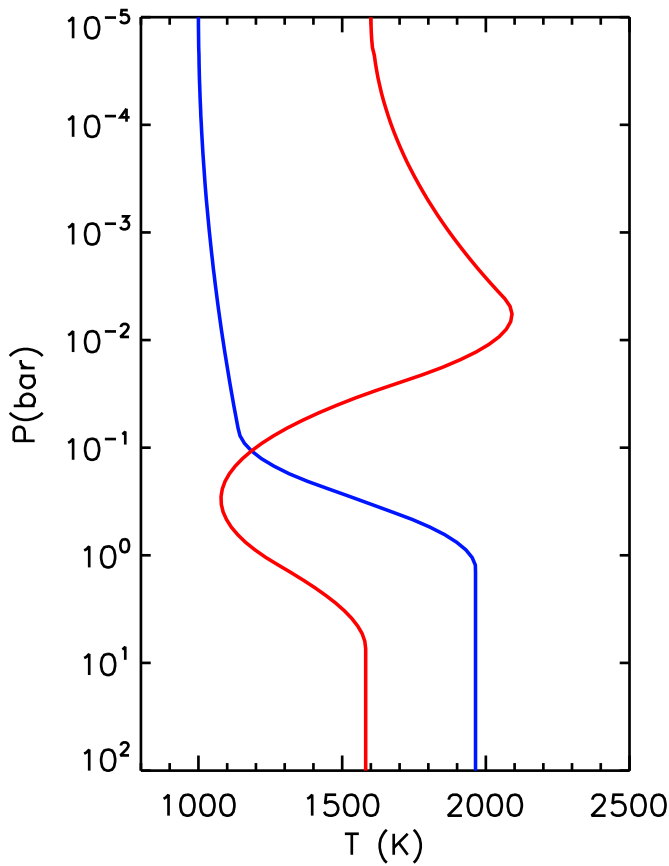}
\caption{Illustration of $P$-$T$ profiles.  The red and blue curves 
  show $P$-$T$ profiles with and without a thermal 
  inversion, respectively. The molecular features due to each of these 
  profiles are shown in Figure~\ref{fig:inv_spec}.}
\label{fig:inv_pt}
\end{figure}

\begin{figure*}[ht]
\centering
\includegraphics[width = \textwidth]{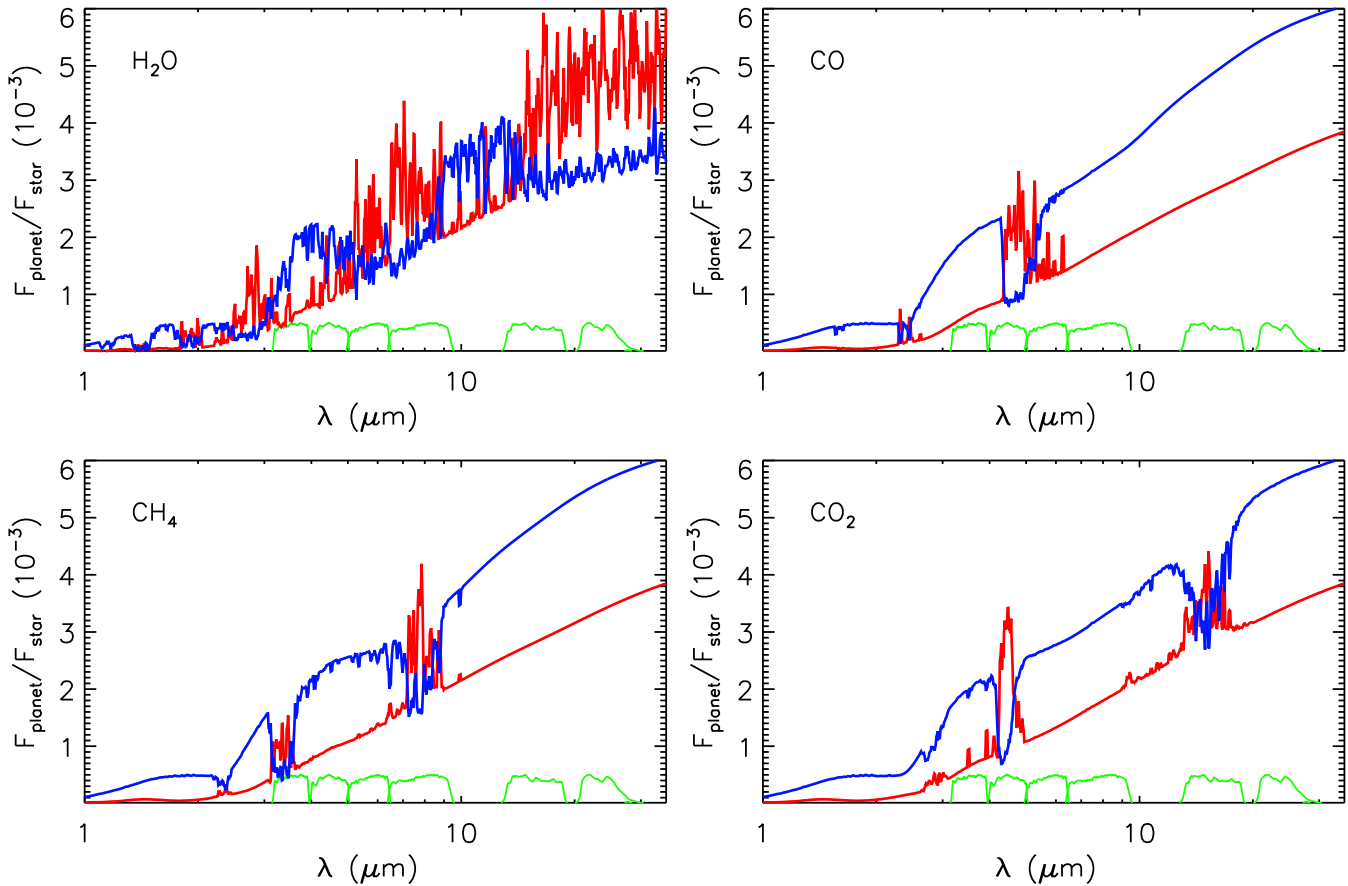}
\caption{Illustration of molecular features. Each panel shows 
  molecular line features due to a single molecule in a hypothetical 
  atmosphere. The stellar and planetary properties are assumed to be those
  of the HD~209458b system. The blue curves show the spectral features of an atmosphere 
  without a thermal inversion, and red curves show features of an atmosphere 
  with a thermal inversion. The corresponding $P$-$T$ profiles are shown in
  Figure~\ref{fig:inv_pt}. The mixing ratios, with respect to H$_2$, in the four panels 
  are assumed to be: H$_2$O = $10^{-4}$, CO = $10^{-3}$, CH$_4$ = $10^{-5}$, 
  CO$_2$ = $5\times10^{-7}$. The continuum in each spectrum is of the black-body 
  of the bottom of the atmosphere with features due to H$_2$-H$_2$ collision
  induced opacities. The green curves show the six {\it Spitzer} band passes at
  3.6, 4.5, 5.8, 8, 16, and 24 $\micron$.}
\label{fig:inv_spec}
\end{figure*}

The thermal inversion causing absorbers of solar system planet
atmospheres do not survive the temperatures of hot
Jupiters. Nevertheless, it has been proposed that thermal inversions
could be formed in the atmospheres of very hot Jupiters due to strong
absorption of incident stellar radiation in the visible by gaseous TiO
and VO (Hubeny et al. 2003; Burrows et al. 2007; Fortney et al. 2008). 
However, Spiegel et al. (2009) suggested that at mbar pressures where thermal inversions 
are required to explain the observations, TiO and VO may not be abundant in the required 
amounts. TiO, being a heavy species, requires a substantial amount of vertical mixing 
to keep it aloft; the $K_{zz}$ required is $10^7 - 10^{11}$ cm$^2$/s, for 0.1 - 10 $\micron$ 
condensate size, if a cold trap is present. Since neither the particle sizes nor $K_{zz}$ 
are known, it is uncertain if TiO might be able to explain thermal inversions. And, VO 
is unlikely to help owing to its lower (solar) abundance and lower visible absorption 
compared to TiO (Spiegel et al. 2009). Zahnle et al. (2009) reported that 
photochemically produced Sulfur compounds, HS and S$_2$, could have high 
UV and visible opacities at the temperatures relevant to hot jupiter atmospheres, 
making them potential candidates for causing thermal inversions in hot jupiters. 
More recently, Knutson et al. (2010) suggested that the presence of a thermal 
inversion could be inversely correlated with the activity level of the host star, 
UV flux from the more active stars potentially destroying inversion-causing absorbers.  
Despite the continuing debate on the inversion-causing absorbers in hot 
jupiter atmospheres, thermal inversions have been invoked typically by either 
adopting a parametric opacity source of unknown nature (Burrows et al. 2008), or 
by parametrizing the temperature profile (see e.g. Madhusudhan \& Seager, 2009). 

\subsection{Review of Observational Inference of Thermal Inversions}
\label{sec:inv-obs}

The inferences of thermal inversions are motivated by observations of dayside 
atmospheres of transiting hot Jupiters using {\it Spitzer} photometry. Observations 
of some hot Jupiters indicate excess emission in some {\it Spitzer} channels over others. 
For example, in the four IRAC observations of HD~209458b reported by Knutson et al.(2008), 
there is indication of excess emission in the 4.5 $\micron$ and 5.8 $\micron$ 
channels. The observations show a markedly higher planet-star flux ratio 
in the 4.5 $\micron$ channel compared to the neighboring 3.6 $\micron$ 
channel. And, the flux ratio in the 5.8 $\micron$ channel is higher 
than that in the adjacent channels at 4.5 $\micron$ and 8 $\micron$. 

The thermal emission spectrum of a planet is influenced by a combination of the 
atmospheric temperature structure and molecular absorption. Let us consider a hypothetical 
hot Jupiter atmosphere with the stellar and planetary properties of HD~209458b, and 
consisting of gaseous H$_2$O, CO, CH$_4$, and CO$_2$, at nominal mixing ratios, 
close to assumption of chemical equilibrium at solar abundances. 
In this particular case, we assume that the molecular species are all well 
mixed in the atmosphere, i.e. uniform volume mixing ratios over the  whole 
atmosphere. Figure~\ref{fig:inv_spec} shows the spectral features of each of the 
molecules in such an atmosphere, for $P$-$T$ profiles with and without a thermal 
inversion (shown in Figure~\ref{fig:inv_pt}). The assumed mixing ratios of the molecules 
are: H$_2$O = $10^{-4}$, CO = $10^{-3}$, CH$_4$ = $10^{-5}$, CO$_2$ = $5\times10^{-7}$. 
As is demonstrated in Figure~\ref{fig:inv_spec}, the atmosphere with a thermal 
inversion gives rise to molecular emission features, whereas the one without 
a thermal inversion has absorption features.  

The excess emission in some {\it Spitzer} channels over others can 
be qualitatively explained by a thermal inversion along with some 
key molecular features. As can be seen from Figure~\ref{fig:inv_spec}, 
H$_2$O has several spectral features in the 3.6 $\micron$, 5.8 $\micron$, 
8 $\micron$, 16 $\micron$, and 24 $\micron$ {\it Spitzer} channels. 
CH$_4$ has strong features almost exclusively in the 3.6 $\micron$ and 8
$\micron$ channels. CO has a strong feature in the 4.5 $\micron$
channel, also contributing to the 5.8 $\micron$ channel. And, CO$_2$
has strong features in the 4.5 $\micron$ and 16 $\micron$ channels. The 
4.5 $\micron$ feature of CO$_2$ is degenerate with the contribution of 
CO in the same channel. The high fluxes in the 4.5 $\micron$ and 
5.8 $\micron$ channels, as seen in HD~209458b for example, can therefore 
be explained simply by having strong emission due to CO and CO$_2$, and only 
moderate emission due to CH$_4$ and H$_2$O (Madhusudhan \& Seager, 2009). 
And, since emission features can form only due to a thermal inversion, 
the observations can be interpreted as suggesting the presence of a thermal 
inversion along with CO and/or CO$_2$.

The molecular species required to explain the observations are physically 
plausible. In chemical equilibrium (Burrows \& Sharp, 1999), CO occurs naturally  
in the temperature range of HD~209458b (see $P$-$T$ profiles in Burrows et al. 
2008, Madhusudhan \& Seager, 2009, Swain et al. 2009). And, CO$_2$ mixing ratios 
up to $\sim 10^{-6}$ are allowed by equilibrium chemistry and/or photochemistry 
(Liang et al. 2003; Zahnle et al. 2009). More generally, hot Jupiters for which
thermal inversions are predicted, are characterized by their 
very hot day-side atmospheres (Fortney et al. 2008), suggesting that dominant 
contribution to the emergent spectra is expected from CO (Burrows \& Sharp, 1999).

Care must be exercised while exploring the space of atmospheric composition in
order to fit the data. Since a high flux in the 4.5 $\micron$ channel
can be explained by CO and/or CO$_2$, it is possible for a fitting
model to infer the lack of CO in the atmosphere, by allowing an implausibly 
high CO$_2$. Such a proposition would likely be implausible given the hot temperatures where
CO is expected to be the dominant form of carbon. The degeneracy
between CO and CO$_2$ can be broken by fitting the model to the 16
$\micron$ {\it Spitzer} IRS photometry, where available.  The 16
$\micron$ channel has dominant contribution due to a strong feature of
CO$_2$ alone. For example, an observation of HD~209458b in the 16 $\micron$ 
channel by Deming (personal communication, 2009) was used in Madhusudhan \&
Seager (2009) to place simultaneous constraints on CO and CO$_2$. The 
same observation is also used in the present work.

\subsection{Alternate Qualitative Interpretation}
\label{sec:inv-alternate}
Is there any conceivable scenario, independent of existing models in
which the observations can be explained without thermal inversions?  A
few qualitative alternatives seem feasible. Let us re-consider the
situation explained in \S~\ref{sec:inv-obs}, where the fluxes in the 
4.5 $\micron$ and 5.8 $\micron$ IRAC channels are higher than those 
in the 3.6 $\micron$ and 8 $\micron$ channels, respectively. Let us now 
investigate if we can explain the same observations with a planet 
atmosphere which has no thermal inversion (the blue model in 
Figure~\ref{fig:inv_pt} and Figure~\ref{fig:inv_spec}). In this case, 
instead of considering the emission features in the 4.5 $\micron$ and 
5.8 $\micron$ channels, one can consider the absorption features in the 
3.6 $\micron$ and 8 $\micron$ channels, to explain the same flux 
differential between adjacent channels. The fitting model would then 
require strong absorption features in the 3.6 $\micron$ and 8 $\micron$ 
channels, and weaker absorption in the 4.5 $\micron$ and 5.8 $\micron$ 
channels. One conceivable solution is provided by methane (CH$_4$) which 
has strong absorption features only in the 3.6 $\micron$ and 8 $\micron$
channels. So, in principle, a high contribution due to CH$_4$, over CO
or CO$_2$, could provide the required absorption signatures. Although
H$_2$O also has features in the 8 $\micron$ channel, a high abundance 
of it may not be desirable since it would lead to absorption in the 
5.8 $\micron$ channel. In presenting this solution, we have exploited 
the degeneracy between the presence or absence of a thermal inversion 
and complementary molecular features.

Despite our simple qualitative explanation of a non-thermal-inversion
fit to observations like those of HD~209458b, whether or not a model 
without inversion fits the data, and is physically plausible, is subject 
to further tests. Firstly, CH$_4$ has stronger absorption in the 8 $\micron$ 
channel than in the 3.6 $\micron$ channel. In other words, for the same
molecular composition of CH$_4$, the absorption in the 8 $\micron$
channel with respect to the 5.8 $\micron$ channel can be deeper than
the absorption in the 3.6 $\micron$ channel with respect to the 4.5
$\micron$ channel. This is contrary to what is required by the IRAC 
observations of HD~209458b (Knutson et al. 2008). 

A fitting model without a thermal inversion must also satisfy the 
constraint of energy balance. Explaining the high flux in the 4.5 
$\micron$ channel with a non-thermal inversion model means that the 
black-body continuum of the spectrum must be at the level of the 
4.5 $\micron$ point or higher. While such a high emergent flux 
balances the incident stellar flux remains to be verified. 
Therefore, it is not certain that a non-inversion configuration which fits 
the data necessarily satisfies the fundamental constraint of energy balance. 
This latter point could, in principle, be obviated by $P$-$T$ profiles steep 
enough to produce deep spectral features in most parts of the spectrum except 
in the 4.5$\micron$ and 5.8 $\micron$ channels. Finally, even if the model fits
and maintains energy balance, it is not clear how such extremely hot
atmospheres can be dominated by CH$_4$ over CO. Future theoretical
work might explain this possibility.

\subsection{A Test for Thermal Inversions}

A rigorous inference of a thermal inversions from a given set of observations 
would involve running a large population of models thoroughly exploring the 
parameter space in search of non-inversion models that fit the data. Not finding 
a statistically significant fit in such an exploration would constitute strong 
evidence in favor of thermal inversions from the data set in question. In a recent 
work, we demonstrated the capability of running $\sim 10^7$ 1-D models on a parameter 
grid with manageable computational resources (Madhusudhan \& Seager, 2009). In the 
same study, we had explored models on predetermined grids in the parameter 
space of models for HD~189733b and HD~209458b. We were then able to 
constrain the extent of possible thermal inversions in HD~209458b that fit the 
{\it Spitzer} broadband photometry. Such a capability was possible because of 
the efficient parametrization of the model temperature structure and molecular 
abundances.

In this study, we combine the model developed in Madhusudhan \& Seager (2009) 
with an efficient parameter space exploration procedure to test the requirement of thermal 
inversions for a select sample of hot Jupiters. We consider four hot Jupiters at different 
levels of irradiation for which observations have been known to be consistent with 
thermal inversions in their respective atmospheres, and for which {\it Spitzer} photometry 
is available in four or more channels. We run $\sim 10^6$ models, with and without thermal 
inversions, for each planet under consideration, and report goodness-of-fit contours 
in the space of atmospheric composition and temperature structure. In what follows, we 
describe our model set-up, the optimization algorithm, and the systems considered in 
this study.

\section{Model and Method}
\label{sec:model}

\subsection{Model}
\label{sec:model-model}

 Our model atmosphere includes a 1D parametric $P$-$T$ profile coupled with line-by-line 
 radiative transfer, hydrostatic equilibrium, and the requirement of 
 energy balance at the top of the atmosphere (Madhusudhan \& Seager, 2009). We consider 100 
 atmospheric layers in the pressure range between $10^{-5} - 100$ bar. 
 The key aspect of our model is the parametrization of the 
 $P$-$T$ profile and the chemical composition, which allows us to 
 run large ensembles of models, exploring the parameter space, in a 
 computationally efficient manner.

 The major difference of our model from traditional atmosphere models
 is in the treatment of energy balance. Our model requires energy
 balance at the top of the atmosphere, instead of an 
 iterative scheme to ensure layer-by-layer radiative (or radiative +
 convective) equilibrium as is done in conventional models. We note that 
 the requirement of layer-by-layer radiative equilibrium in a 1-D model is 
 not strictly physical since complex hydrodynamics flows in highly irradiated 
 hot Jupiter atmospheres can alter the temperature structure away from 
 radiative equilibrium (Showman et al. 2009). The global energy balance, 
 e.g., at the top of the atmosphere, however, is a strict requirement. For a given 
 set of model parameters, we require that the net energy output
 at the top of the atmosphere is less than or equal to the net energy
 input due to the incident stellar flux; a deficit indicates energy
 redistributed to the night-side. Models where the emergent flux is
 greater than the incident flux are discarded (see Madhusudhan \&
 Seager, 2009). By running a large number of ($\sim 10^7$) models in
 the parameter space, and discarding those that did not satisfy energy
 balance, we were left with a population of models that satisfied energy
 balance.

 We parameterize the chemical abundances of the molecular species by 
 considering deviations over chemical equilibrium (explained at 
 length in Madhusudhan and Seager, 2009). For each molecule under 
 consideration, we compute its mixing ratio in a layer of the atmosphere 
 by multiplying a parametric factor to the mixing ratio that would be 
 expected under thermochemical equilibrium with solar abundances (TE$_\odot$).
 The parametric factor for a given molecule is constant over the 
 entire atmosphere, i.e the mixing ratio profile of the molecule over 
 the entire atmosphere would be shifted relative to that obtained from 
 TE$_\odot$ by the constant factor. We use this treatment for H$_2$O, CO, 
 and CH$_4$. For CO$_2$, we perturb over a uniform mixing ratio of 
 $10^{-6}$ (1 ppmv or $10^{-6}$ is just a reference; it is the approximate mixing ratio 
 of CO$_2$ expected from TE$_\odot$ and/or photochemistry, for 5 $\times$ solar metallicity 
 and  $T \sim 2000 K$; Zahnle et al. 2009, Liang et al. 2003). Thus corresponding to the four prominent 
 molecules, we have four parameters: $f_{\rm H_2O}$, $f_{\rm CO}$, $f_{\rm CH_4}$, and $f_{\rm CO_2}$. 
 We reiterate that $f_X$ is not the absolute mixing ratio of ``X''. It 
 is the ratio between the mixing ratio of ``X'' and the mixing ratio under TE$_\odot$; 
 except for CO$_2$, for which it is with respect to $10^{-6}$. Our models 
 also include NH$_3$, fixed at the TE$_\odot$ value, and TiO and VO at solar 
 abundances. Additionally, we include H$_2$-H$_2$ collison induced cross-sections,
 which are a source of continuum opacity. Our H$_2$O, CH$_4$, CO and NH$_3$ 
 molecular line data are from Freedman et al. (2008), and references therein. 
 Our CO$_2$ data are from Freedman (personal communication) and Rothman et al. 
 (2005). And, we obtain the H$_2$-H$_2$ collision-induced opacities from 
 Borysow et al. (1997), and Borysow (2002). We use a Kurucz model for the stellar 
 spectrum (Castelli \& Kurucz, 2004).
 
 In the current work, we have made one significant change to our approach in 
 Madhusudhan \& Seager (2009). Previously we had run models on a 
 predetermined grid, chosen based on some model independent arguments
 outlined in that work. While we were able to run a large population
 (tens of millions) of models in that approach, the grid resolution was 
 still coarse and sampled evenly over evidently unnecessary regions of 
 the parameter space. In this approach, we run the models using a more 
 efficient parameter space exploration procedure, allowing us to sample 
 the desired error surfaces at much higher resolution.

\subsection{Parameter Space Exploration}
\label{sec:model-mcmc}
Our primary requirement in this work is to be able to explore 
the model parameter space at fine resolution. Even a single 
plausible model without thermal inversions, in a million models, 
would still be evidence against the requirement of thermal 
inversion by a given set of observations. 

We use the Markov chain Monte Carlo (MCMC) method to explore the 
parameter space of models without thermal inversions. 
The MCMC method is a Bayesian parameter estimation algorithm 
which allows the calculation of posterior probability distributions 
of the model parameters conditional to a given set of observations.
An extensive body of literature exists on the applications of 
MCMC for parameter estimation (Gilks et al. 1998; Tegmark et al. 2004; 
Ford et al. 2005). The MCMC method allows an efficient means 
of exploring the parameter space in search of a global solution, with 
very fine sampling in the allowed range of parameter values. In this work, 
however, the observations are always less than the number of parameters, 
i.e there is no unique solution. However, it is still possible to explore 
the parameter space and find contours in the error surface of some measure 
of fit. We, therefore, use the MCMC method with a Metropolis-Hastings
scheme within the Gibbs sampler, for fine sampling of the model parameter space.

Our model described in \S~\ref{sec:model-model} above has ten free 
parameters. Six parameters concern the $P$-$T$ profile:
T$_0$, P$_1$, P$_2$, P$_3$, $\alpha_1$, and $\alpha_2$ (Madhusudhan \& Seager, 2009). 
And, four parameters correspond to the departures of molecular abundances from the 
reference abundances described in \S~\ref{sec:model-model}: $f_{\rm H_2O}$, $f_{\rm CO}$, 
$f_{\rm CH_4}$, and $f_{\rm CO_2}$. 

We define some physically motivated boundaries in the parameter space explored by the 
Markov chain. We impose the constraint of global energy balance by 
restricting $\eta$ to [0.0,1.0], where, $\eta = (1-A)(1-f_r)$ is the ratio of emergent flux 
output on the day-side to incident stellar flux input on the day-side, weighted 
appropriately (Madhusudhan \& Seager, 2009). Here, $A$ is the Bond Albedo and $f_r$ 
is the day-night energy redistribution. And, we impose some nominal boundaries on the 
temperatures and departures from equilibrium chemistry. We explore a wide range of deviations 
from chemical equilibrium (Burrows \& Sharp, 1999), empirically selected so as to be general 
enough. For models without thermal inversions, we set the 
boundaries as $-9 < \log(f_{\rm H_2O}) < 9$, $-9 < \log(f_{\rm CO}) < 5$,  $-5 < \log(f_{\rm CH_4}) < 10$, 
$-5 < \log(f_{\rm CO_2}) < 4$. The limits are similar for models with thermal inversions, except the 
lower boundaries for $\log(f_{\rm H_2O})$ and $\log(f_{\rm CO})$, which are set at -4 and -5, 
respectively. We report all those models which have the overall elemental C/H and O/H abundances 
within the broad range of $(10^{-2} - 10^2)~\times$ solar (Anders \& Grevesse, 1989; 
Burrows \& Sharp, 1999; but c.f.~Asplund \& Grevesse, 2005; Allende-Prieto et al. 2002). For the 
temperature structure, the constraint of no thermal inversion is imposed by requiring that $P_1 \ge P_2$. The ``fit'' parameters for the MCMC are T$_0$, $\log$(P$_1$), $\log$(P$_2$), $\log$(P$_3$), $\alpha_1$, $\alpha_2$, $\log$($f_{\rm H_2O}$), $\log$($f_{\rm CO}$), $\log$($f_{\rm CH_4}$), and $\log$($f_{\rm CO_2}$). 
We consider uniform priors in all the parameters. For each system under consideration, 
we run one chain of $10^6$ links for models with thermal inversion and one for models 
without thermal inversion. Our parametric $P-T$ profile provides a simple means to 
demarcate between inversion and non-inversion models. The condition for the $P$-$T$ 
profile to have no thermal inversion is $P_2 \leq P_1 \leq P_3$. And, that to have 
a thermal inversion is $P_1 \leq P_2 \leq P_3$.

\begin{figure*}[ht]
\centering
\includegraphics[width = \textwidth]{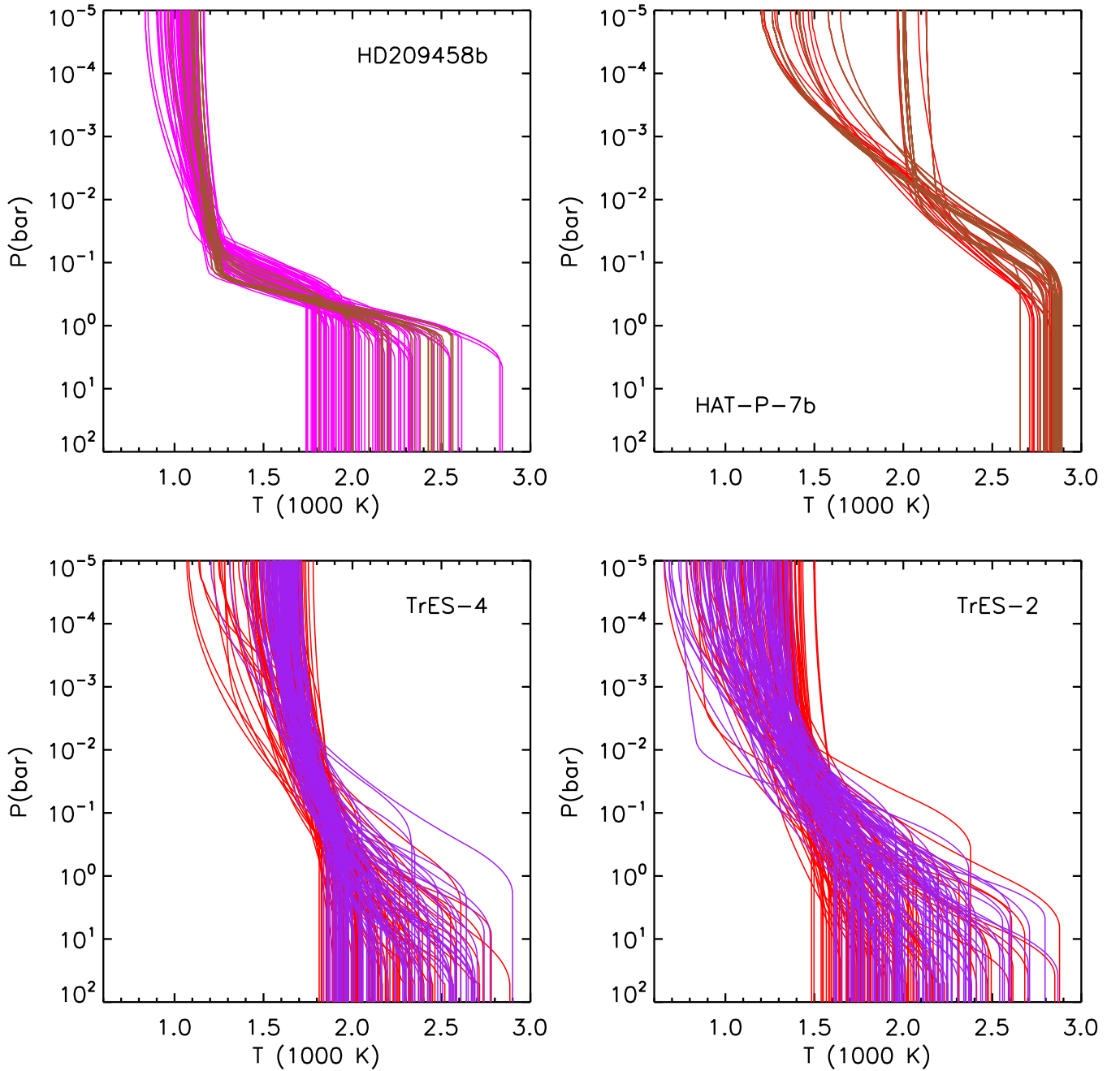}
\caption{Pressure-Temperatures ($P$-$T$) profiles for the four
  systems. Each panel shows the $P$-$T$ profiles with no thermal
  inversions that explain the observations at different levels of fit. 
  Best-fitting profiles with thermal inversions for each of these systems 
  have been reported in the literature referred in the text. 
  For HD~209458b, the profiles in magenta correspond to models that
  fit the observations with $2 < \xi^2 \le 2.25$; the best-fit non-inversion 
  model has $\xi^2$ = 2.04. For HAT-P-7b, the red profiles correspond to 
  models fitting the observations to within $1.6 < \xi^2 \le 2$; the best-fit 
  model had $\xi^2 = 1.65$. The brown profiles for both HD~209458b and 
  HAT-P-7b are 30 profiles that fit best, shown for illustration. For TrES-4 
  and TrES-2 the red profiles correspond to models that fit within $1 \le \xi^2 \le 2$,
  and the purple profiles fit to within $\xi ^2 < 1$; only 100 randomly 
  chosen profiles from each category are shown, for clarity.}
\label{fig:pt}
\end{figure*}

\begin{figure*}[]
\centering
\includegraphics[width = 5.0in, height = 7in]{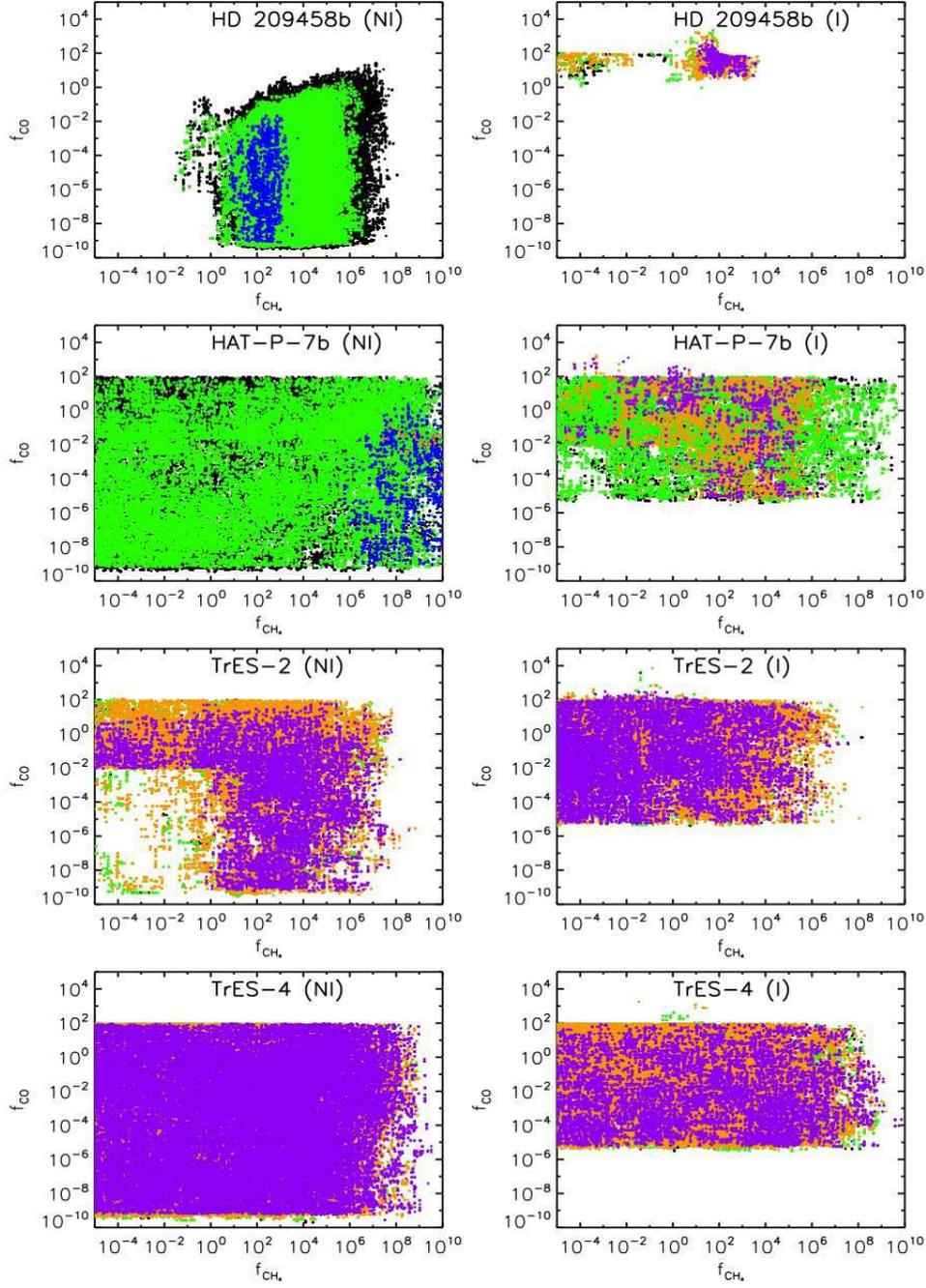}
\caption{Departures from chemical equilibrium of CH$_4$ and CO. The dots indicate 
the regions in the space of CO and CH$_4$ mixing ratios explored by 
the MCMC chain (see Section~\ref{sec:model-mcmc}); each dot is a model realization.
All models with C/H and O/H within $(10^{-2} - 10^2)$ $\times$ solar are shown. The 
boundaries in the composition space are described in section~\ref{sec:model-mcmc}.
For each planet, f$_{\rm CO}$ and f$_{\rm CH_4}$ are the departures in 
the mixing ratios of CO and CH$_4$ from those corresponding to thermochemical 
equilibrium with solar abundances (TE$_\odot$) for the same temperature structure
(see Section~\ref{sec:model-model}). For example, f$_{\rm CO}$ = 1 implies a 
CO concentration that is at TE$_\odot$. The left (right) panel for each system 
shows constraints on models without (with) thermal inversions. Non-inversion 
and inversion models are labelled with (NI) and (I), respectively. The purple, orange, 
green and black colors correspond to $\xi ^2 \leq 1$, $1 \le \xi^2 \le 2$, 
$2 \le \xi^2 \le 3$, and $3 \le \xi^2 \le 4$, respectively. The blue dots for 
non-inversion models in HD~209458b and HAT-P-7b correspond to 
$2.0 < \xi^2 \le 2.25$; the best fitting model for HD~209458b has a $\xi^2 = 2.04$, 
and that for HAT-P-7b has  $\xi^2 = 1.65$.}
\label{fig:c_all}
\end{figure*}

\begin{figure*}[]
\centering
\includegraphics[width = 5.0in, height = 7in]{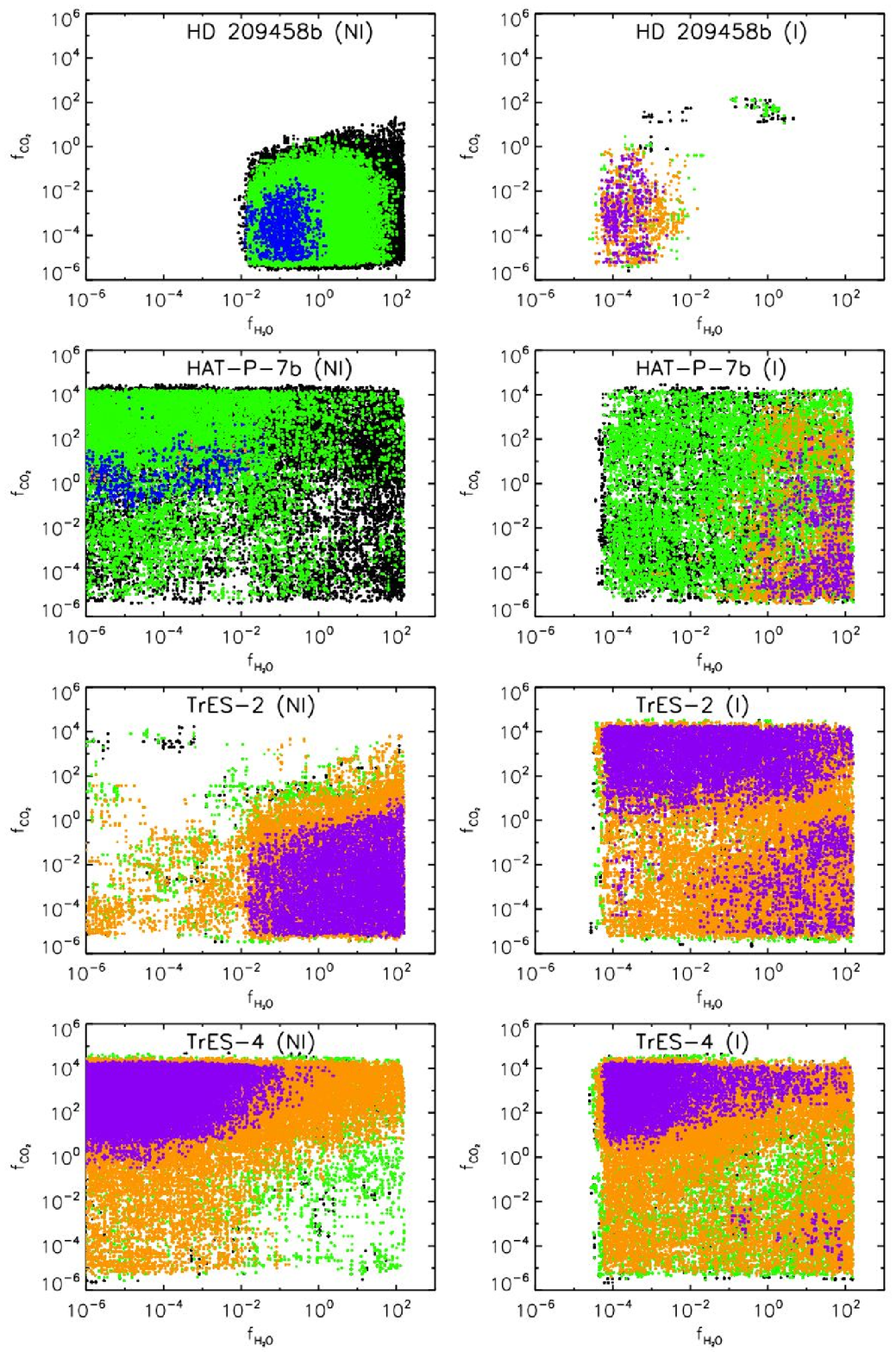}
\caption{Departures from chemical equilibrium of H$_2$O and CO$_2$. 
For each planet, f$_{\rm H_2O}$ is the departure in H$_2$O mixing ratio from 
that corresponding to TE$_\odot$ (see Figure~\ref{fig:c_all}), and f$_{\rm CO_2}$ 
is the departure in CO$_2$ mixing ratio from a constant value of $10^{-6}$ 
(see Section~\ref{sec:model-model}). The description of panels and colors is 
identical to that in Figure~\ref{fig:c_all}. All models with C/H and O/H within 
$(10^{-2} - 10^2)$ $\times$ solar are shown. The boundaries in the composition 
space are described in section~\ref{sec:model-mcmc}.}
\label{fig:c_all_1}
\end{figure*}

\subsection{Quantitative Measure of Fit}

Central to our analysis is the definition of what constitutes a
``fit'' to the data. We can only report to what extent a model fits
the data relative to the ``observational" uncertainties, i.e within the 1 $\sigma$ error
bars, or 1.5 $\sigma$ errors, and so on. Given that the number of
broadband observations are typically less than the number of model
parameters, we cannot report a formal fit with confidence levels.
Nevertheless, we evaluate our models based on the $\xi^2$ statistic, defined as
$\chi^2/N_{obs}$ (Madhusudhan \& Seager, 2009):
\begin{equation}
\xi^2 = \frac{1}{N_{obs}} \sum_{i = 1}^{N_{obs}} \bigg(\frac{f_{i,model} - f_{i,obs}}
{\sigma_{i,obs}} \bigg)^2,
\label{eq:xi}
\end{equation}
where, $f_{i,model}$ and $f_{i,obs}$ are the model and observed flux ratios, respectively, 
and $\sigma_{i,obs}$ is the 1 $\sigma$ measurement uncertainty. $N_{obs}$ is the number
of observations.

For each system, we report the best value of $\xi^2$ we find with a
non-inversion model. We also present the range in parameter space
which fit the observations at different levels of $\xi^2$, for
example, $\xi^2 \le 1$, $\xi^2 \le 2$, or higher, as applicable. In
this framework, $\xi^2 \le 1$ means the model fits the observations to
within the 1~$\sigma$ error bars on average. And, a $\xi^2 \le 2$
means a fit to within $\sqrt 2 = 1.4$~$\sigma$ of the error bars on
average, and so on.

\begin{deluxetable*}{l c c c c c c}
\tablewidth{0pt} 
\tabletypesize{\scriptsize}
\tablecaption{Constraints on the atmospheric properties for models without thermal inversions}
\tablehead{\colhead{}   &   \colhead{HD~209458b\tablenotemark{a}}   &   \colhead{HAT-P-7b\tablenotemark{a}}   &   \multicolumn{2}{c}{TrES-4}   &   \multicolumn{2}{c}{TrES-2} \\\\
  \cline{2-2} \cline{3-2} \cline{4-5} \cline{6-7} \\
  \colhead{}   &   \colhead{$2 < \xi^2 \le 2.25$}   &   \colhead{$1.5 < \xi^2 \le 2$}   &   \colhead{$\xi^2 \le 1$}   &   \colhead{$\xi^2 \le 2$}     &   \colhead{$\xi^2 \le 1$}   &   \colhead{$\xi^2 \le 2$}}
  \startdata  
H$_2$O\tablenotemark{b}   &   $10^{-5} - 2 \times 10^{-3}$               &   $10^{-12} - 2\times 10^{-5}$   &   $10^{-12} - 2 \times 10^{-3}$   &   $10^{-13} - 0.15$                &   $8 \times 10^{-6} - 0.16$     &   $10^{-12} - 0.16$ \\
CO                                            &   $10^{-12} - 10^{-5}$                            &   $3\times10^{-8} - 10^{-4}$      & $10^{-13} - 0.06$                       &   $10^{-13} - 0.07$                  &   $10^{-13} - 0.04$                  &   $10^{-13} - 0.07$ \\
CH$_4$                                   &   $6\times10^{-6} - 2\times 10^{-4}$   &   $2\times10^{-3} - 0.05$           &   $10^{-15} - 0.07$                       &   $10^{-15} - 0.07$                &   $10^{-15} - 0.07$                  &   $10^{-15} - 0.07$ \\
CO$_2$                                   &   $10^{-11} - 4 \times 10^{-8}$             &   $9 \times 10^{-7} - 9 \times 10^{-5}$    &   $3\times10^{-7}  - 0.02$           &   $10^{-12} - 0.03$ &   $10^{-11} - 10^{-5}$             &   $10^{-12} - 7 \times 10^{-3}$ \\
C/O                                           &   $0.03 - 0.82$                                        &   $15 - 2 \times 10^{3}$              &   $0.44 - 4 \times 10^3$              &   $10^{-4} - 4 \times 10^3$   &   $5\times10^{-5} - 4 \times 10^{3}$   &   $4\times10^{-5} - 4\times10^{3}$ \\
$\eta$\tablenotemark{c}       &   $0.5 - 1.0$                                             &   $0.7-1.0$                                 &   $0.37 - 1.0$                                 &   $0.33 - 1.0$                            &   $0.36 - 1.0$                            &   $0.29 - 1.0$ 
\enddata
\tablenotetext{a}{For HD~209458b and HAT-P-7b, the best-fit non-inversion model has $\xi^2$ of 2.04 and 1.65, respectively, and 
  hence the reported ranges of $\xi^2$. See text for details.}
\tablenotetext{b}{The molecular mixing ratios are quoted as ratios by number with respect to molecular hydrogen.}
\tablenotetext{c}{$\eta = (1-f_r)(1-A)$, where, $f_r$ is the day-night redistribution, and A is the bond albedo. (1-$\eta$) gives the maximum day-night redistribution allowed by the model, i.e assuming zero albedo.}
\label{tab:tab1_noinv}
\end{deluxetable*}

\begin{deluxetable*}{l c c c c c c} 
\tablewidth{0pt} 
\tabletypesize{\scriptsize}
\tablecaption{Constraints on the atmospheric properties for models with thermal inversions}
\tablehead{\colhead{}   &   \multicolumn{2}{c}{HD~209458b}   &   \multicolumn{2}{c}{HAT-P-7b}   &   \colhead{TrES-4\tablenotemark{a}}   &   \colhead{TrES-2\tablenotemark{a}} \\\\
  \cline{2-3} \cline{4-5} \cline{6-6} \cline{7-7} \\
  \colhead{}   &   \colhead{$\xi^2 \le 1$}   &   \colhead{$\xi^2 \le 2$}  &  \colhead{$\xi^2 \le 1$}   &   \colhead{$\xi^2 \le 2$}  &   \colhead{$\xi^2 \le 1$}  &   \colhead{$\xi^2 \le 1$}}
  \startdata  
H$_2$O\tablenotemark{b}   &   $5\times10^{-8} - 3 \times 10^{-6}$  &   $4\times10^{-8} - 10^{-5}$      &   $10^{-4} - 0.16$        &   $3\times10^{-7} - 0.16$      &   $5\times10^{-8} - 0.13$  &   $5\times10^{-8} - 0.16$   \\
CO                                            &   $3\times10^{-4} - 0.06$                      &   $3\times10^{-4} - 0.07$           &   $3 \times 10^{-9} - 0.07$        &   $10^{-9} - 0.07$    &   $4\times10^{-9} - 0.07$                &   $10^{-9} - 0.07$                \\
CH$_4$                                   &   $9\times10^{-4} - 0.06$                      &   $10^{-11} - 0.06$                     &   $10^{-14} - 0.07$      &   $10^{-15} - 0.07$                 &   $10^{-15} - 0.07$               &   $10^{-14} - 0.07$              \\
CO$_2$                                   &   $10^{-11} - 5 \times 10^{-7}$             &   $10^{-12} - 8 \times 10^{-7}$ &   $10^{-11} - 2 \times 10^{-4}$ &   $10^{-12} - 0.01$  &   $10^{-11}  - 0.02$             &   $10^{-11} - 0.02$              \\
C/O                                           &   $1 - 146$                                               &   $1 - 105$                                   &   $5\times10^{-5} - 53$            &   $5\times10^{-5} - 207$&   $4\times10^{-4} - 2 \times 10^3$&   $6\times10^{-5} - 10^{3}$   \\
$\eta$\tablenotemark{c}       &   $0.27 - 0.60$                                         &   $0.25 - 0.84$                            &   $0.49 - 0.83$              &   $0.41- 0.85$                         &   $0.36 - 1.0$                       &   $0.23 - 1.0$  
\enddata

\tablenotetext{a}{For TrES-2 and TrES-4, the constraints at the  $\xi^2 = 2$ level are almost identical to those at the $\xi^2 = 1$ level, and hence we do not report them here.}
\tablenotetext{b}{The molecular mixing ratios are quoted as ratios by number with respect to molecular hydrogen.}
\tablenotetext{c}{$\eta = (1-f_r)(1-A)$, where, $f_r$ is the day-night redistribution, and A is the bond albedo. (1-$\eta$) gives the 
  maximum day-night redistribution allowed by the model, i.e assuming zero albedo.}
\label{tab:tab1_inv}
\end{deluxetable*}

\section{Results}

In this section, we report the constraints on the chemical compositions of 
the day-side atmospheres of the planets in our study. For each planet, we 
present the range of composition, temperature structure, and day-night 
energy redistribution required by the best-fit models with no thermal inversions. 
We also present the constraints on the composition and day-night energy 
redistribution of models with thermal inversions for each system. 

\subsection{HD~209458b}
\label{sec-hd209}

We consider planet-star flux contrasts of HD~209458b in six channels of
{\it Spitzer} broadband photometry. The data include four IRAC 
observations reported by Knutson et al. (2008), and observations 
in the 16 $\micron$ IRS channel and the 24 $\micron$ MIPS channel, 
by Deming (personal communication) and Deming et al.(2005), 
respectively. We focus on these observations which were reported in the 
literature as suggestive of a thermal inversion in HD~209458b, especially 
the four IRAC observations. 

Our results indicate that HD~209458b is a likely candidate to host a thermal 
inversion in its day-side atmosphere. However, whether or not HD~209458b 
actually has a thermal inversion depends on the level of fit, and the physical 
plausibility of the fitting models one is willing to consider. Figure~\ref{fig:pt} 
shows populations of pressure-temperature ($P$-$T$) profiles with no 
thermal inversions which fit the observations at different levels of $\xi^2$. 
The corresponding constraints on the atmospheric composition are shown 
in Table~\ref{tab:tab1_noinv}, and the constraints for inversion models are 
shown in Table~\ref{tab:tab1_inv}. Figure~\ref{fig:c_all} and Figure~\ref{fig:c_all_1} 
show departures of molecular species from TE$_\odot$, for models of HD~209458b 
with and without thermal inversions. 

The observations require a thermal inversion in the atmosphere of HD~209458b 
at the $\xi^2 = 2$ level. The best fitting model with no thermal inversion 
has a $\xi^2$ of 2.04, implying a fit at 1.43 $\sigma$ (i.e $\sqrt{2.04}$) 
of the observations, on average. And, even at this level of fit, the models 
show substantial departures from thermochemical equilibrium assuming 
solar abundances (TE$_\odot$). Figure~\ref{fig:c_all} shows the departures 
in the mixing ratios of CO and CH$_4$ from TE$_\odot$ at different levels of 
fit. At the $2.0 < \xi^2 < 2.25$ surface (shown in blue dots for the non-inversion 
case), it can be seen that non-inversion models of HD~209458b require a 
depletion of CO of at least $10^{-2}$ times TE$_\odot$ (the departures shown 
in Figure~\ref{fig:c_all} are in fraction with respect to TE$_\odot$). This low a 
mixing ratio of CO can, in principle, be achieved by having similarly low 
abundances of C and O relative to solar. However, the simultaneous requirement 
of an overabundance of CH$_4$, is hard to explain. Thus, non-inversion models 
fitting the observations at the 1.5 $\sigma$ errors seem physically implausible. 

The observations can be explained by physically plausible non-inversion models  
at the $\xi^2 \sim 3$ level, meaning a fit at the 1.7 $\sigma$ observational errors. As shown in 
Figure~\ref{fig:c_all}, the $\xi^2 = 3$ level (region with green dots) allows for 
non-inversion models which have CO within a factor of $\sim 10$
from TE$_\odot$. Such small a factor can potentially be explained either by just having 
different C/H and O/H abundances or due to non-equilibrium processes 
(Cooper \& Showman, 2006; Zahnle et al. 2009; Line et al. 2010; Madhusudhan \& Seager, 2010). 

\begin{figure*}[ht]
\centering
\includegraphics[width = \textwidth]{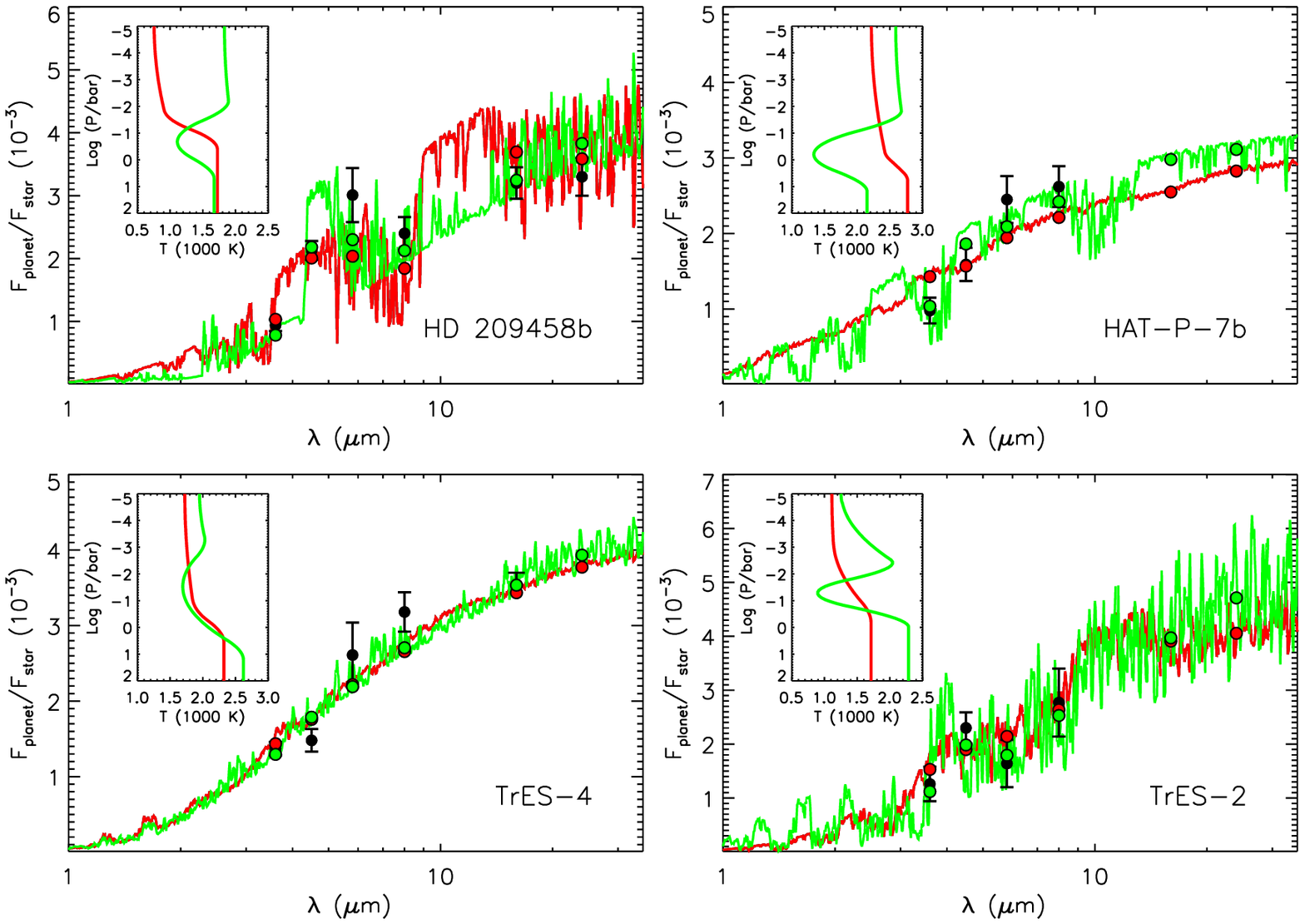}
\caption{Sample model spectra for each system. Two model spectra, 
corresponding to models with and without a thermal inversion, are presented 
for each system. The models represent a balance between degree of fit and physical 
plausibility. The corresponding $P$-$T$ profiles are shown in the insets. The atmospheric 
compositions and day-night redistribution corresponding to each model 
are shown in Table~\ref{tab:spectra}.  In each panel, the black circles with 
error bars (and the upper-limit for TrES-4) are the available {\it Spitzer} 
observations (see text for details). The red and green circles are the channel 
integrated model spectra in the six {\it Spitzer} photometric channels, 
corresponding to the red and green curves, respectively.}
\label{fig:spectra}
\end{figure*}

The observations can be fit to within $\xi^2 = 1$ by models with thermal 
inversions, as has been demonstrated previously by Madhusudhan \& Seager (2009). 
The molecular mixing ratios for inversion models as constrained by the observations 
are shown in Table~\ref{tab:tab1_inv}. Figures~\ref{fig:c_all} and \ref{fig:c_all_1} show 
the deviations from TE$_\odot$ required by the inversion models. We find that the 
best-fitting models, within $\xi^2 = 1$, allow compositions with CO and CH$_4$ 
deviant from TE$_\odot$ by a factor of $\sim 10$ and higher. However, if we 
consider the $\xi^2 = 2$ surface, the observations can be fit with inversion 
models containing close to TE values of CO and 
CH$_4$, with C/H and O/H abundances only slightly enhanced over solar.
One potential problem in the inversion scenario, however, is the requirement 
of low H$_2$O by the best-fit models. The low observed flux in the 3 $\micron$ 
and 24 $\micron$ channels requires low H$_2$O, which is contrary to the 
high H$_2$O requirement imposed by the high observed flux in the 5.8 $\micron$ 
channel. This problem has been previously discussed in the literature 
(Deming et al. 2005, Seager et al. 2005, Madhusudhan \& Seager, 2009), 
and is a subject for future studies.

Two best-fit model spectra, with and without a thermal inversion, for HD~209458b 
are shown in Figure~\ref{fig:spectra}, along with the $P$-$T$ profiles. 
The corresponding atmospheric composition and day-night redistribution are 
shown in Table~\ref{tab:spectra}. As can be seen from Figure~\ref{fig:spectra}, 
the dominant source of error for the non-inversion model comes from the high 
flux in the 5.8 $\micron$ IRAC observation. While most of the observations can 
be fit at the $\sim 1 \sigma$ error bars, the 5.8 $\micron$ is fit only at $\sim 2.5 \sigma$. 
The dominant source of opacity in this channel comes from H$_2$O. In the 
non-inversion scenario, a low water content could potentially explain the lack 
of absorption in this channel but will over predict the flux in the 24 $\micron$ channel. 

\subsection{HAT-P-7b}

HAT-P-7b is one of the hottest transiting hot Jupiters known. 
Being the hottest of our sample of planets, it is also the most 
expected hot Jupiter in our sample to host a thermal inversion. We use 
the day-side observations of HAT-P-7b reported in four channels 
of {\it Spitzer} IRAC photometry by Christiansen et al. (2010). The 
constraints on the molecular abundances, and departures from 
equilibrium, for the inversion and non-inversion models are shown 
in Table~\ref{tab:tab1_noinv}, Table~\ref{tab:tab1_inv}, Figure~\ref{fig:c_all}, 
and  Figure~\ref{fig:c_all_1}.

Our results confirm previous findings that the inversions models fit 
the observations of HAT-P-7b better than models without thermal 
inversions. As shown in Figure~\ref{fig:c_all}, inversion models can fit 
the observations to within the 1-$\sigma$ errors (i.e. $\xi^2 \leq 1$) for 
a wide range of methane and CO concentrations, including those close 
to TE$_\odot$ values. On the other hand, the best fitting non-inversion 
model has a $\xi^2 = 1.65$, indicating a fit at 1.3-$\sigma$ errors. Even 
then, the best fitting non-inversion models of HAT-P-7b shown in 
Figure~\ref{fig:c_all} (blue dots) require methane abundances that are 
over five orders of magnitude greater than the TE$_\odot$ values, which  
are seemingly implausible.  

The observations of HAT-P-7b can plausibly be explained without a thermal 
inversion at the $\xi^2 = 3$ level, i.e fits at 1.7-$\sigma$ errors (green dots). 
At this level, models with methane and CO concentrations only marginally 
deviant from TE$_\odot$ values can explain the data. However, such a degree 
of fit may not be statistically representative of the true nature of the planet 
atmosphere. A sample of non-inversion temperature profiles at different levels 
of fit is shown in Figure~\ref{fig:pt}. The $P$-$T$ profiles at $\xi^2 \leq 2$ are shown 
in Figure~\ref{fig:pt}. The brown profiles show the 30 best-fit $P$-$T$ profiles below $\xi^2 = 2$. 

Two model spectra, corresponding to models with and without a thermal inversion, 
are shown in Figure~\ref{fig:spectra}, along with the corresponding temperature 
profiles.  The model parameters for each case are shown in Table~\ref{tab:spectra}. 
As shown in Figure~\ref{fig:spectra}, the non-inversion model is unable to 
fit the low flux at 3.6 $\micron$ and the high flux at 8 $\micron$, simultaneously. 
The reason for this behavior is because both channels have absorption features 
due to methane. A low (high) methane concentration in a non-inversion model causes 
high (low) observed flux in both the channels, contrary to what is observed. 
On the other hand, the inversion model easily explains all the observations. 
The low flux in the 3.6 micron channel is explained by the temperature decreasing 
outward in the lower layers, and the high flux at 8 microns is explained by 
the high temperatures, due to the thermal inversion, in the higher layers where 
the contribution to the 8 $\micron$ channel peaks.

The day-night redistribution is well-constrained by the data. At the
$\xi^2 = 2$ level, the constraint on $\eta = (1-A)(1-f_r)$ is 0.7 -
1.0, for non-inversion models. This range allows for a maximum redistribution 
of 0.26, assuming zero albedo, and hence implies relatively inefficient 
advection of energy to the night side. This finding is consistent with the low
redistribution of an inversion model reported in Christiansen et
al. (2010), and the finding of low redistribution in the visible light
curve of {\it Kepler}. The inversion models on the other hand allow for 
a wider range of $\eta = 0.41 - 0.85$ at the $\xi^2 \leq 2$ level, allowing 
for models with efficient day-night redistribution and those otherwise.

In principle, the allowed parameter space of the models can be further
restricted by the observation in the visible (centered at 0.63 $\micron$) 
obtained by the {\it Kepler} space telescope (Borucki et al. 2009). 
However, modeling the visible flux in the {\it Kepler} bandpass introduces
another five free parameters in terms of the dominant opacity sources
in the visible - the atmospheric concentrations of TiO, VO, Na, K, and a
prescription for scattering. These five parameters, which are 
largely decoupled from the opacity sources in the IR, allow a large 
degree of flexibility in fitting the one {\it Kepler} observation. 
By exploring a preliminary range of values for the five parameters, we 
do find that some of our best-fit non-inversion models for the {\it Spitzer} 
data are also able to fit the {\it Kepler} point. In an earlier work, we found 
that the {\it Kepler} point could also be fit with models with thermal inversions 
(Christiansen et al, 2010). 

\begin{deluxetable*}{l c c c c c c c}
\tablewidth{0pt} 
\tabletypesize{\scriptsize}
\tablecaption{Chemical compositions and day-night redistribution of sample spectra\tablenotemark{a}}
\tablehead{\colhead{} & \colhead{H$_2$O} & \colhead{CO} & \colhead{CH$_4$} & \colhead{CO$_2$} & \colhead{C/O} & \colhead{$\eta$} & \colhead{$\xi^2$}}
\startdata  
HD~209458b  \\
I   & $  6\times 10^{-06} $ (0.006) & $  3\times 10^{-02} $ (50)  & $  4\times 10^{-08} $ (0.001) & $ 6\times 10^{-09} $ (0.006) & 1.00 & 0.38 & 1.60\\
NI  & $  4\times 10^{-05} $ (0.04)  & $  8\times 10^{-06} $ (0.01)& $  9\times 10^{-06} $ (0.8)   & $ 5\times 10^{-11} $ ($5\times 10^{-05}$) & 0.36 & 0.56 & 2.75\\
\hline \\
HAT-P-7b \\
I   & $  7\times 10^{-03} $ (7) & $  6\times 10^{-04} $ (0.8) & $  8\times 10^{-07} $ (0.1) & $ 1\times 10^{-05} $ (10) & 0.07 & 0.59 & 0.89\\
NI  & $  2\times 10^{-05} $ (0.02) & $  5\times 10^{-04} $ (0.63) & $  2\times 10^{-11} $ (0.01) & $ 8\times 10^{-05} $ (80) & 0.85 & 0.63 & 3.0\\
\hline \\
TrES-4 \\ 
I   & $  8\times 10^{-03} $ (8) & $  4\times 10^{-04} $ (0.6) & $  5\times 10^{-08} $ (1) & $ 2\times 10^{-08} $ (0.02) & 0.05 & 0.40 & 1.76\\
NI  & $  5\times 10^{-03} $ (5)  &  $  4\times 10^{-04} $ (0.6) & $  2\times 10^{-9} $ (0.2) & $ 2\times 10^{-05} $ (23) & 0.09 & 0.41 & 1.80\\
\hline \\
TrES-2 \\
I   & $  2\times 10^{-03} $ (2) & $  1\times 10^{-03} $ (2) & $  1\times 10^{-05} $ (0.3) & $ 1\times 10^{-08} $ (0.01) & 0.44 & 0.72 & 0.54\\
NI  & $  1\times 10^{-03} $ (1) & $  4\times 10^{-04} $ (0.6) & $  2\times 10^{-08} $ (0.7) & $ 2\times 10^{-08} $ (0.02) & 0.23 & 0.46 & 0.98
\enddata
\tablenotetext{a}{The two rows for each system correspond to the two model spectra for each system
  presented in Figure~\ref{fig:spectra}. ``I"  and ``NI" correspond to the models with and without a thermal 
  inversion, respectively, in Figure~\ref{fig:spectra}. For each molecule, the mixing ratio averaged over all
   the layers is reported, along with the deviation from TE$_{\odot}$ shown in parentheses.}
\label{tab:spectra}
\end{deluxetable*}

\subsection{TrES-4}

TrES-4 is the second hottest planet in our sample, after HAT-P-7b, and 
is highly favored to host a thermal inversion, on theoretical grounds 
(Fortney et al. 2008). We use the five {\it Spitzer} photometric observations 
of Knutson et al. (2009), which have been previously reported as evidence 
for a thermal inversion in TrES-4. 

We find that existing observations of TrES-4 can be explained almost equally 
well by models with and without thermal inversions, contrary to previous findings. 
Our fits for each case are better than $\xi^2 = 1$, i.e within the 1-$\sigma$ 
error bars. The non-inversion $P$-$T$ profiles fitting the data at the 
$\xi^2 \leq 1$ and $\xi^2 \leq 2$ levels are shown in Figure~\ref{fig:pt}. 

As shown in Figure~\ref{fig:c_all}, non-inversion models can fit the observations 
of TrES-4 to within $\xi^2 \leq 1$, for a wide range of CO and CH$_4$ abundances, 
including the values at TE$_\odot$, represented by  $f_{\rm CO} = 1$ and $f_{\rm CH_4} = 1$. 
However, we find that the best-fitting models (within $\xi^2 \leq 1$) without inversions require a high 
concentration of CO$_2$, $\gtrsim 10^{-4}$, for TE$_\odot$ concentrations of the remaining 
molecules like H$_2$O (e.g. Figure~\ref{fig:c_all_1}). This requirement arises from the 
non-detection of planet flux observed in the 16 $\micron$ channel. Since the dominant 
contribution to this channel comes from the CO$_2$ feature at 15 $\micron$, a non-detection 
of flux indicates substantial absorption due to CO$_2$ and hence the high CO$_2$ requirement. 
At the high temperatures of TrES-4, a CO$_2$ concentration of $10^{-4}$ is not feasible via  
equilibrium chemistry at solar abundances. However, CO$_2$ concentrations as high as 
$\sim10^{-4}$ are, in principle, feasible for a high metallicity, over $30 \times$ solar, (Zahnle et al. 2009;  Madhusudhan \& Seager, 2010).
  
Non-inversion model fits at the $\xi^2 \leq 2$ level, on the other hand, do allow for a 
very plausible set of chemical compositions. Figure~\ref{fig:spectra} shows a model 
spectrum without a thermal inversion (in red) and having a chemical composition close to 
TE$_\odot$ (Table~\ref{tab:spectra} shows the composition). Also shown for reference is a 
model with a thermal inversion. The constraints on the molecular abundances are shown in Table~\ref{tab:tab1_noinv} and Table~\ref{tab:tab1_inv}, for models without and with thermal inversions, 
respectively. We conclude that the observations place almost no constraints on the molecular 
concentrations in the atmosphere of TrES-4, in either scenario. While the wide range in allowed 
chemical compositions is seemingly implausible, it does allow for a significant population of 
non-inversion models that are physically plausible (see \ref{sec:plausible}). We find this evidence 
enough to conclude that there is no sure sign of a thermal inversion in TrES-4, given current data. 

\subsection{TrES-2}

Our results show that the observations of TrES-2 can be explained to 
a high degree of fit by models both with and without thermal inversions. 
This general conclusion is similar to that of O'Donovan et al. (2010), where 
we first reported that the data could be fit by models with and without thermal 
inversions. In that study our non-inversion models fitting the data 
required a CO abundance that was lower than TE$_\odot$ by about two 
orders of magnitude, whereas the best-fit inversion models  allowed TE$_\odot$ 
composition. However, with the new parameter exploration routine, in the present 
study we have been able to explore regions of parameter space well beyond what 
we could pursue in O'Donovan et al. (2010). While our present results for inversion 
models agree with our previous findings, our results for non-inversion models go 
beyond our findings in O'Donovan et al. (2010). 

The best-fit solutions with no thermal inversions in the present study span a 
wide range in chemical composition, including that of TE$_\odot$. Our results show 
that the $P$-$T$ profile is mostly unconstrained by the data, resulting in a large region 
of parameter space that can explain the observations even at the 
$\xi^2 = 1$ level. Figure~\ref{fig:pt} shows the non-inversion $P$-$T$ 
profiles corresponding to $\xi^2 \leq 1$ and $\xi^2 \leq 2$. 

The constraints on the atmospheric composition and day-night redistribution are 
shown in Tables~\ref{tab:tab1_noinv} and \ref{tab:tab1_inv}, and in Figures~\ref{fig:c_all} 
and \ref{fig:c_all_1} . For each scenario, i.e., with or without a thermal inversion, there is 
a wide range in molecular composition that can explain the data, including that of 
thermochemical equilibrium with solar abundances (i.e TE$_\odot$). The composition 
in each scenario is practically unconstrained even at the $\xi^2 \leq 1$ level. It follows 
that the C/O ratio is also unconstrained by the data, spanning a rather unphysical 
range of 5 $\times 10^{-5}$ - 4 $\times 10^{3}$ even at the $\xi^2 = 1$ level, for the 
non-inversion models, for example. The large range in C/O 
is a consequence of allowing the molecular mixing ratios to vary arbitrarily. 
Nevertheless, it does indicate that best-fit solutions with no thermal inversions can 
be found at the $\xi^2 \leq 1$ level with very plausible molecular concentrations (see
\S~\ref{sec:plausible}), and both carbon dominated and oxygen dominated
atmospheres are allowed by the observations. Similarly, the day-night
redistribution is also unconstrained by the data, at the $\xi^2 = 1$ level.

Based on the existing {\it Spitzer} IRAC observations, therefore, our results show 
that a thermal inversion cannot be inferred in the day-side atmosphere of TrES-2.
The weak constraints on the atmosphere of TrES-2 is evident from the
data. The flux ratio in the 4.5 $\micron$ IRAC channel is noticeably 
higher than that in the 3.6 $\micron$ channel, hinting at a possible
emission feature due to a thermal inversion. However, the flux ratio
in the 5.8 $\micron$ channel is noticeably lower than that in the 8
$\micron$ channel allowing for H$_2$O absorption, and hence the lack
of a thermal inversion. Thus, while a model with thermal inversion can
explain the data, a thermal inversion is not required by existing IRAC
observations of TrES-2. Two best-fit models for TrES-2, with and without 
a thermal inversion, are shown in Figure~\ref{fig:spectra}, and the 
corresponding compositions are shown in Table~\ref{tab:spectra}. The 
models for both the cases shown here span a rather plausible range of 
chemical compositions. 

\section{Summary and Discussion}
\label{sec:discussion}

We have investigated the question of whether thermal inversions can be 
robustly inferred from existing {\it Spitzer} photometric observations of thermal 
emission from hot Jupiter atmospheres. We addressed this objective by 
thoroughly exploring the parameter space of models with and without 
thermal inversions for a sample of four hot Jupiters. We considered 
four systems which have {\it Spitzer} observations at four or more wavelengths, 
and are highly irradiated, so that they are theoretically favored to host thermal 
inversions (Hubeny et al. 2003; Fortney et al. 2008): HD~209458b, HAT-P-7b, TrES-4, and TrES-2. 
Furthermore, the observations considered have also been previously reported 
to be consistent with thermal inversions in the corresponding systems, albeit 
less robustly. In this work, we addressed to what level of statistical significance and 
physical plausibility thermal inversions can be inferred in each of these systems. 

Our primary finding is that a detailed exploration of the model parameter space 
is necessary to make robust inferences of thermal inversions in exoplanetary 
atmospheres. We find that the observations of TrES-4 and TrES-2 can be explained by 
models with and without thermal inversions, and with physically plausible 
chemical compositions, at the $\xi^2 \le 2$ and $\xi^2 \le 1$ levels, respectively. This finding 
is in contrast to the findings of Knutson et al. (2009) and Spiegel \& Burrows (2010) who, based on 
forward models of Burrows et al. (2008), have suggested the requirement of thermal inversion 
from the same data sets. The difference in conclusions results from the different 
modeling schemes; the forward models of Burrows et al (2008) assume layer-by-layer radiative 
equilibrium and chemical equilibrium, and a parametrization for day-night energy 
redistribution. On the other hand, for a given dataset, we explore the space of temperature 
profile and composition, without the requirement of layer-by-layer radiative equilibrium 
or chemical equilibrium, but still imposing the strict constraint of global energy balance. 
We note that our best-fitting inversion solutions for TrES-2 and TrES-4 do include a wide range of 
profiles with thermal inversions, which likely encompass the profiles of Knutson et al. (2008) and 
Spiegel \& Burrows (2010). Another interesting result is that if TrES-4 indeed does not host a thermal 
inversion, the best-fitting non-inversion solutions require a large CO$_2$ mixing ratio ($\ge 10^{-5}$), 
which might be an indication of enhanced metallicity in TrES-4 (Zahnle et al. 2009; 
Madhusudhan \& Seager, 2010).

For HD~209458b and HAT-P-7b, we find that the observations 
cannot be explained without thermal inversions to within a $\xi^2 < 3$, i.e to 
within $1.7\sigma$ observational uncertainties, for any plausible composition. Any better 
fit would require substantial enhancements in methane and depletion of CO, 
which is implausible at the high temperatures in the systems considered. 
Our inference of thermal inversions in HD~209458b and HAT-P-7b is consistent with 
previous findings of Burrows et al. (2008), Knutson et al. (2008), Christiansen et al. (2010), 
and Spiegel \& Burrows (2010). Our results show that a detailed exploration of the model 
parameter space and an accurate assessment of the observational errors is essential to 
robustly infer thermal inversions based on existing photometric observations. For example, 
if one considers the 2-$\sigma$ error bars on the data, thermal inversions may not be required 
even for compositions in chemical equilibrium.

\subsection{Thermal Inversions or Not?}

Whether or not the observations considered in this work can be 
explained without a thermal inversion depends on what level of 
fit, and physical plausibility of models, one is willing to consider. 
If we do not consider the physical plausibility of the best-fit models, 
the observations of all the four hot Jupiters can be explained without 
thermal inversions to within the 1.5 $\sigma$ error bars, 
i.e $\xi^2 \le 2.25$. On the other hand, if we consider only models 
fitting within the 1$\sigma$ error bars, and/or enforce arguments 
of physical plausibility (see \S~\ref{sec:plausible} below), HAT-P-7b 
and HD~209458b emerge as likely to host thermal inversions.
The inference of a thermal inversion is, therefore, very sensitive to the 
reported observational uncertainties.

The inference of a thermal inversion can also be sensitive to one data point 
over others. In HD~209458b, for instance, several of the six observations can 
be fit at the $\sim 1 \sigma$ error bars by non-inversion models (e.g., 
Figure~\ref{fig:spectra}). A large contribution to the $\xi^2$ comes 
predominantly from the 5.8 $\micron$ point which is fit only at greater 
than 2$\sigma$. Thus, for these models it is only the 5.8 $\micron$ IRAC point 
which guides any inference we make about thermal inversions. Similarly, for the 
model spectra shown for HAT-P-7b, the dominant contribution to  $\xi^2$ comes 
from the 8 $\micron$ IRAC point. Therefore, any inference of thermal inversions 
can be highly sensitive to the reported observational uncertainties in a single 
channel, which varies on a case by case basis. 

At the level of current observations, our results potentially deviate from theory. 
Given that the atmospheres of TrES-4 and TrES-2 can be explained 
by models without thermal inversions, it is possible that these systems do 
not host thermal inversions. If that happens to be the case, the results are 
in contrast to theoretical predictions, as both TrES-4 and TrES-2 have higher 
levels of incident star flux as compared to HD~209458b (Fortney et al. 2008), 
and hence are more likely to host thermal inversions. Nevertheless, since 
the observations for these systems are consistent with models both with and 
without thermal inversions, the only conclusion is that it is too early to claim 
thermal inversions in these systems, contrary to some previous studies.

\subsection{Plausibility of Models}
\label{sec:plausible}

An important point concerns the physical plausibility of 
non-inversion models fitting the observations. As shown in 
Figure~\ref{fig:c_all}, the best-fit non-inversion models for HD~209458b 
and HAT-P-7b require substantial enhancement of CH$_4$ as compared to 
CO. However, as explained in \S~\ref{sec:inv-alternate}, in the very hot 
atmospheres of these planets, CO is expected to be the dominant 
molecule, based on atmospheric chemistry. We currently do not 
have a physically plausible explanation for such CH$_4$ enhancement 
at the expense of CO in a very-hot atmosphere. 

Our best-fit models explore an unrestricted  range of atmospheric compositions.
In trying to conduct an unbiased exploration of the parameter space, 
we have allowed for all the molecules to vary over a large 
range of values, that might be seemingly unphysical. For example, 
in TrES-2 the $\xi^2 = 1$ limits extend to mixing ratios as 
high as 0.1 for H$_2$O, CO and CH$_4$. Such high mixing ratios 
indicate extreme metallicities that are too high to be plausible, although 
not impossible. For reference, solar abundances have number fractions 
of C and O at $3.3 \times 10^{-4}$ and $7.8 \times 10^{-4}$, respectively 
(Anders \& Grevesse, 1989; used in equilibrium chemistry calculations of 
Burrows \& Sharp, 1999; but c.f. Allende Prieto et al.,~2002; Asplund et al.,~2005, 
for recent values which are lower by a factor of $\sim$ 2). Similarly, for HAT-P-7b and 
TrES-2 the C/O ratios span many orders of magnitude reaching 
values as high as $10^4$, which are also manifestly unphysical. For instance, the 
solar C/O ratio is $\sim 0.5$. However, the best-fitting models, with and/or 
without inversions, for all the systems do allow C/O ratios in the plausible range of 0.1 - 1. 

Finally, we have not explored the realm of drastically inhomogeneous models - 
those models where the mixing ratio of a molecule could be very different in 
different layers of the atmosphere. It is understandable that photochemistry 
and vertical mixing can deviate molecular mixing ratios away from equilibrium 
(e.g. Line et al. 2010; Madhusudhan \& Seager, 2010). We expect that our 
prescription for molecular species, which is parametrized as deviations from 
chemical equilibrium, spans the space of possible deviations. However, we have 
not considered arbitrarily populating different layers of the atmosphere with different 
species. Adhoc filling of the layers with specific molecules might allow a large, albeit 
unphysical, degree of freedom in fitting the observations. 

\subsection{Future Observations to Resolve the Degeneracy}

In this work, we have addressed the apparent degeneracy between atmospheric 
composition and thermal inversions in hot Jupiter atmospheres. The large 
number of model parameters allows the freedom to fit the limited observations 
of some atmospheres in any scenario, i.e with or without inversions; although, 
at different levels of fit. Future developments in observations and theory are 
needed to break the apparent degeneracies. Theoretical efforts are needed to 
put limits of physical plausibility on the atmospheric composition and temperature 
structure. Such limits, for example, might exclude many of the non-inversion 
models that fit the observations considered in this work. 

New observations are important for better constraints on thermal inversions. 
In the near future, multiple observations of thermal emission from transiting hot Jupiters 
in the near-IR, from ground and with {\it HST}, along with existing {\it Spitzer} data 
can help constrain models to a good extent, on a case-by-case basis. For instance, the two 
model spectra of HAT-P-7b in Figure~\ref{fig:spectra} show only modest differences in 
the {\it Spitzer} bandpasses. However, the spectra are markedly different in the near-IR, 
especially in the continua between the molecular features, which can potentially be 
observed from ground, e.g., in the $J$, $H$, and $K$ bands. Additional constraints 
can be also placed by observations within the molecular features, which are possible 
with space-based observations (e.g. with {\it HST}), except for molecules like methane 
where ground-based observations might also be feasible. Near-IR observations have 
been reported for a few systems to date, not particularly constraining thermal inversions 
(e.g. Swain et al. 2009; Croll et al. 2010). However, multi-band near-IR photometry and/or 
spectroscopy can prove to be a rich resource for targeted searches for thermal inversions 
in exoplanetary atmospheres. Targets can be selected based on constraints from already 
existing {\it Spitzer} observations, irradiation levels, and sensitivity of a given instrument to 
the planet-star flux contrasts. The several near-IR bandpasses mentioned above are 
currently ripe for this purpose. In the long run, high resolution spectra with the {\it James 
Webb Space Telescope} will have the sensitivity to conclusively identify the presence of 
thermal inversions based on spectrally resolved emission features.
 
An important point concerns the opportunity to observe in the two IRAC channels on {\it Warm Spitzer}. 
While it is true that robust inferences of thermal inversions could not be made 
even with four {\it Spitzer} points in several known cases, a large difference between 
the 3.6 $\micron$ and 4.5 $\micron$ channels can place stringent constraints on the 
existence of thermal inversions. A very large flux excess in the 4.5 $\micron$ channel 
over the 3.6 $\micron$ channel is highly indicative of a thermal inversion; although, 
it also depends on the irradiation level of the planet which governs the blackbody 
continuum. A large excess in the 3.6 $\micron$ channel, on the other hand, is almost 
a sure sign of no thermal inversion, as in the cases of HD~189733b (Madhusudhan \& Seager, 2009) 
and GJ~436b (Stevenson et al. 2010; Madhusudhan \& Seager, 2010). Thus, 
observations of hot Jupiters with {\it Warm Spitzer} would likely be able to identify the 
extreme cases of systems with or without thermal inversions. 

Tremendous progress has been made in the last decade in our understanding 
of exoplanetary atmospheres. At the same time, recent and current observations 
allow us a chance to revisit previous interpretations made with limited 
observations. It is now upon us to evaluate all the theoretical options and 
observational uncertainties so as to determine a framework in which to 
interpret observations. Judicious target-selection and efficient planning of 
future observations, from ground and from space, will be critical to characterizing the 
atmospheres of the growing number of transiting exoplanets.

\acknowledgements {We thank Heather Knutson and Jonathan Fortney for helpful 
  discussions. This work is based on published observations made with
  the {\it Spitzer Space Telescope}, which is operated by the Jet
  Propulsion Laboratory, California Institute of Technology under a
  contract with NASA. Support for this work was provided by NASA
  through an award issued by JPL/Caltech.}



\vspace{1mm}

\begin{thebibliography}{}
\bibitem[Allende(2002)]{Allende:02} Allende Prieto, C., Lambert, D. L. \& Asplund, M. 2002, \apj, 573, L137
\bibitem[Anders(1989)]{Anders:89} Anders, E., \& Grevesse, N. 1989, Geochimica et Cosmochimica Acta, 53, 197
\bibitem[Asplund(2005)]{Asplund:05}Asplund, M., Grevesse, N., \& Sauval, A. J. 2005, in ASP Conf. Ser. 336, Cosmic Abundances as Records of Stellar Evolution and Nucleosynthesis, ed. T. G. Barnes, III \& F. N. Bash (San Francisco, CA: ASP), 25
\bibitem[Barman(2005)]{Barman:05} Barman, T. S., Hauschildt, P. H., \& Allard, F. 2005, \apj, 632, 1132
\bibitem[Borucki(2009)]{Borucki:09} Borucki, W. J., et al. 2009, Science, 325, 709
\bibitem[Borysow(1997)]{Borysow:97} Borysow, A.,  Jorgensen, U. G., \& Zheng, C. 1997, \aa, 324, 185. 
\bibitem[Borysow(2002)]{Borysow:02} Borysow, A. 2002, \aa, 390, 779. 
\bibitem[Burrows(1999)]{Burrows:99} Burrows, A. \& Sharp, C. M. 1999, \apj, 512, 843 
\bibitem[Burrows(2007)]{Burrows:07}Burrows, A., Hubeny, I., Budaj, J., Knutson, H. A., \&  Charbonneau, D. 2007, \apj, 668, L171
\bibitem[Burrows(2008)]{Burrows:08} Burrows, A., Budaj, J. \& Hubeny, I. 2008, \apj, 678, 1436
\bibitem[Castelli(2004)]{Castelli:04}Castelli, F. \& Kurucz, R. L. 2004, arXiv:astro-ph/0405087v1 (ftp://ftp.stsci.edu/cdbs/grid/ck04models/).
\bibitem[Chamberlain(1978)]{Chamberlain:78} Chamberlain, J. W. 1978, Theory of Planetary Atmospheres, Academic Press, Inc., New York, NY.
\bibitem[Christiansen(2010)]{Christiansen:10}Christiansen, J., et al. 2010, \apj, 710, 97 
\bibitem[Cooper(2006)]{Cooper:06}Cooper, C. S. \& Showman, A. P. 2006, \apj, 649, 1048 
\bibitem[Croll(2010)]{Croll:10} Croll, B., Albert, L., Lafreniere, D., Jayawardhana, R. \& Fortney, J. J., \apj, 717, 1084
\bibitem[Deming et al.(2005)]{Deming:05} Deming, D., Seager, S., Richardson, L.\ J., \& Harrington, J. 2005, \nat, 434, 740
\bibitem[Ford(2005)]{Ford:05} Ford, E. 2005, \apj, 129, 1706
\bibitem[Fortney(2006)]{Fortney:06}Fortney, J. J., Saumon, D., Marley, M. S., Lodders, K. \& Freedman, R. S. 2006, \apj, 642, 495
\bibitem[Fortney(2008)]{Fortney:08}Fortney, J. J., Lodders, K., Marley, M. S. \& Freedman, R. S. 2008, \apj, 678, 1419
\bibitem[Freedman(2008)] {Freedman:08} Freedman, R. S., Marley, M. S. \& Lodders, K. 2008, \apj S, 174, 504.
\bibitem[Fressin(2010)]{Fressin:10} Fressin, F., et al. 2010, \apj, 711, 374 
\bibitem[Gilks(1998)]{Gilks:98}Gilks, W. R., Richardson, S., \& Spiegelhalter, D. J. 1998., Markov chain Monte Carlo in practice, Chapman \& Hall/CRC
\bibitem[Hubeny(2003)]{Hubeny:03} Hubeny, I., Burrows, A., \& Sudarsky, D. 2003, \apj, 594, 1011
\bibitem[Knutson(2008)]{Knutson:08}Knutson, H. A., Charbonneau, D., Allen, L. E., Burrows, A. \& Megeath, S. T. 2008, \apj, 673, 526.
\bibitem[Knutson(2009)]{Knutson:09}Knutson, H. A., Charbonneau, D., Burrows, A., O'Donovan, F. T. \& Mandushev, G. 2009, \apj, 691, 866.
\bibitem[Knutson(2010)]{Knutson:10}Knutson, H. A., Howard, A. W. \& Isaacson, H. 2010, \apj, 720, 1569
\bibitem[Liang(2003)]{Liang:03}Liang, M-C, Parkinson, C. D., Lee, A. Y.-T., Yung, Y. L., \& Seager, S. 2003, \apj, 596, L247.
\bibitem[Line(2010)]{Line:10} Line, M. R, Liang, M-C, \& Yung, Y. L. 2010,  \apj, 717, 496
\bibitem[Machalek(2009)]{Machalek:09} Machalek, P., McCullough, P. R., Burrows, A., Burke, C. J., Hora, J. L., \& Johns-Krull, C. M. 2009, \apj, 701, 51
\bibitem[MS(2009)]{MS:09}Madhusudhan, N. \&  Seager, S. 2009, 707, 24
\bibitem[MS(2010)]{MS:10}Madhusudhan, N. \&  Seager, S. 2010, (arXiv:1004.5121)
\bibitem[Donovan(2010)]{Donovan:10}O'Donovan, F. T., et al. 2010, \apj, 710, 1551
\bibitem[Rothman(2005)]{Rothman:05} Rothman, L. S., et al. 2005, J. Quant. Spec. \& Rad. Transfer, 96, 139
\bibitem[Seager(2005)]{Seager:05} Seager, S., Richardson, L. J., Hansen, B. M. S., Menou, K., Cho, J. Y.-K. \& Deming, D. 2005, \apj, 632, 1122
\bibitem[Showman(2009)]{Showman:09} Showman, A. P., Fortney, J. J., Lian, Y., Marley, M. S., Freedman, R. S., Knutson, H. A. \& Charbonneau, D. 2009, \apj, 699, 564. 
\bibitem[Spiegel(2009)]{Spiegel:09}	Spiegel, D. S., Silverio, K., \& Burrows, A. 2009, \apj, 699, 1487
\bibitem[Spiegel(2010)]{Spiegel:10}Spiegel, D. S. \& Burrows, A. 2010, \apj, 722, 871
\bibitem[Stevenson(2010)]{Stevenson:10} Stevenson, K. et al. 2010. \nat,  464, 1161
\bibitem[Swain(2009)]{Swain:09} Swain, M. R., et al. 2009, \apj, 704, 1616  
\bibitem[Tegmark(2004)]{Tegmark:04} Tegmark, et. al. 2004, Phys. Rev. D, 69, 103501
\bibitem[Zahnle(2009)]{Zahnle:09}Zahnle, K., Marley, M. S., Lodders, K., \& Fortney, J. J. 2009, \apj, 701, L20
\end{thebibliography}
\end{document}